# Tables of crystallographic properties of double antisymmetry space groups

**Mantao Huang[a], Brian K. VanLeeuwen[a], Daniel B. Litvin[b] and Venkatraman Gopalan[a]*** 

[a]Department of Materials Science and Engineering, The Pennsylvania State University, University Park, PA, 16801, USA, and [b]Department of Physics, The Eberly College of Science, The Pennsylvania State University, Penn State Berks, P.O. Box 7009, Reading, PA, 19610, USA

Correspondence email: vgopalan@psu.edu

**Synopsis**

Tables of crystallographic properties of double antisymmetry space groups are presented.

**Abstract**

Tables of crystallographic properties of double antisymmetry space groups, including symmetry element diagrams, general position diagrams, and positions, with multiplicities, site symmetries, coordinates, spin and roto vectors are presented.

## 1. Introduction

Double antisymmetry space groups are symmetry groups that describe rotation-reversal and time-reversal symmetry in crystals. These groups were introduced to generalize the symmetry classification of rigid static rotations in crystals (Gopalan & Litvin, 2011). There are two independent anti-identities representing time inversion and roto inversion, which are labelled 1' and 1* respectively. In addition, there is an anti-identity that is the product of the two independent ones, labelled 1'* (colors seen online only). The colors used to distinguish the identity and three anti-identities are listed in Table 1.

**Table 1** Identity and the three anti-identities of double antisymmetry space groups

| Identity or anti-identity | Color | Properties |
|:---:|:---:|:---:|
| 1 | Black | does not reverse time, and does not reverse rotation |
| 1' | Red | reverses time, and does not reverse rotation |
| 1* | Blue | does not reverse time, and reverses rotation |
| 1'* | Green | reverses time, and reverses rotation |

A full list of the 17,803 proper affine classes of double antisymmetry space groups was presented by VanLeeuwen et al. (VanLeeuwen *et al.*, 2013). These classes are referred to as the 17,803 double antisymmetry space group types or the 17,803 double antisymmetry space groups. For each set of groups belonging to the same double antisymmetry space group type, one representative is chosen and listed (VanLeeuwen *et al.*, 2013). The tables presented in the present work give the crystallographic properties of these representatives.[1] The structure of the tables is similar to the tables found in the International Tables for Crystallography Volumes A (2006) and E (2010), and in tables of magnetic space groups (Litvin, 2013). The details of the format and content of the tables are discussed in Section 2.

---

[1] An electronic document *Double Antisymmetry Space Group Listings* containing these tables is available from http://sites.psu.edu/gopalan/research/symmetry/.

No. 10977    $P(1,1^*,1)mm'a^*$    $Pmmn$
$a, 2b+\frac{1}{2}, c$

SG. 51    $mm'm1^*$    Orthorhombic

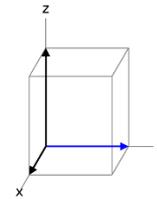

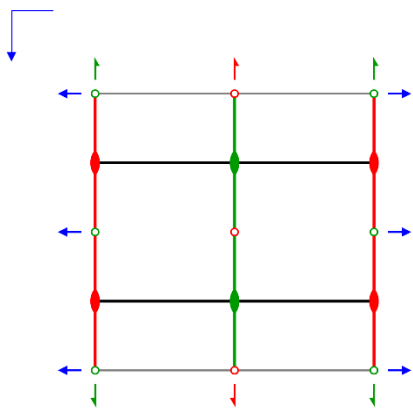

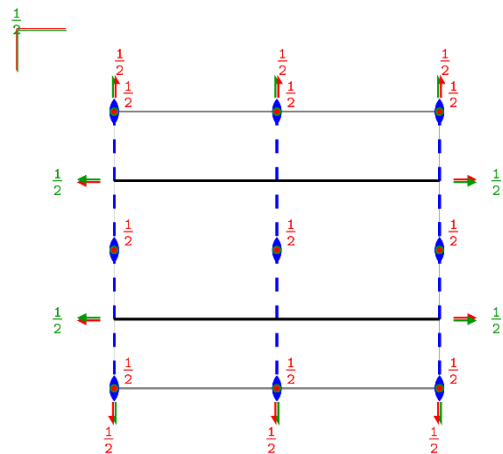

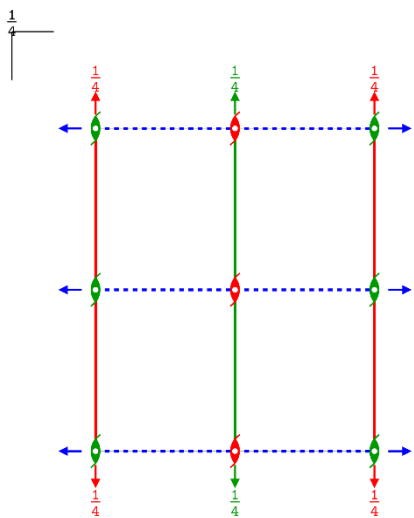

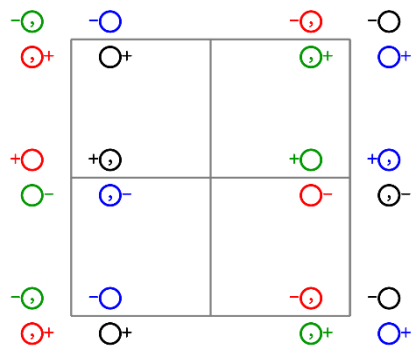

**Symmetry Operations**

(1) 1
  $(1|000)$

(2) $2'\ \frac{1}{4}, 0, z$
  $(2_z|\frac{1}{2}00)'$

(3) $2^*\ 0, y, 0$
  $(2_y|000)^*$

(4) $2\left(\frac{1}{2},0,0\right)'^*\ x, 0, 0$
  $(2_x|\frac{1}{2}00)'^*$

(5) $\bar{1}'^*\ 0, 0, 0$
  $(\bar{1}|000)'^*$

(6) $a^*\ x, y, 0$
  $(m_z|\frac{1}{2}00)^*$

(7) $m'\ x, 0, z$
  $(m_y|000)'$

(8) $m\ \frac{1}{4}, y, z$
  $(m_x|\frac{1}{2}00)$

| Positions | | | Coordinates | | |
|---|---|---|---|---|---|
| 8 | l | 1 | $\{x,y,z\}[a,b,c][d,e,f]$ | | $\{\bar{x}+\frac{1}{2},\bar{y},z\}[\bar{a},\bar{b},c][\bar{d},\bar{e},f]$ |
| | | | $\{\bar{x},y,\bar{z}\}[\bar{a},b,\bar{c}][\bar{d},e,\bar{f}]$ | | $\{x+\frac{1}{2},\bar{y},\bar{z}\}[a,\bar{b},\bar{c}][d,\bar{e},\bar{f}]$ |
| | | | $\{\bar{x},\bar{y},\bar{z}\}[a,b,c][d,e,f]$ | | $\{x+\frac{1}{2},y,\bar{z}\}[\bar{a},\bar{b},c][\bar{d},\bar{e},f]$ |
| | | | $\{x,\bar{y},z\}[\bar{a},b,\bar{c}][\bar{d},e,\bar{f}]$ | | $\{\bar{x}+\frac{1}{2},y,z\}[a,\bar{b},\bar{c}][d,\bar{e},\bar{f}]$ |
| 4 | k | $m..$ | $\{\frac{1}{4},y,z\}[a,0,0][d,0,0]$ | | $\{\frac{1}{4},\bar{y},z\}[\bar{a},0,0][\bar{d},0,0]$ |
| | | | $\{\frac{3}{4},y,\bar{z}\}[\bar{a},0,0][\bar{d},0,0]$ | | $\{\frac{3}{4},\bar{y},\bar{z}\}[a,0,0][d,0,0]$ |
| 4 | j | $.m'^*.$ | $\{x,\frac{1}{2},z\}[a,0,c][d,0,f]$ | | $\{\bar{x}+\frac{1}{2},\frac{1}{2},z\}[\bar{a},0,c][\bar{d},0,f]$ |
| | | | $\{\bar{x},\frac{1}{2},\bar{z}\}[\bar{a},0,\bar{c}][\bar{d},0,\bar{f}]$ | | $\{x+\frac{1}{2},\frac{1}{2},\bar{z}\}[a,0,\bar{c}][d,0,\bar{f}]$ |
| 4 | i | $.m'.$ | $\{x,0,z\}[a,0,c][0,e,0]$ | | $\{\bar{x}+\frac{1}{2},0,z\}[\bar{a},0,c][0,\bar{e},0]$ |
| | | | $\{\bar{x},0,\bar{z}\}[\bar{a},0,\bar{c}][0,e,0]$ | | $\{x+\frac{1}{2},0,\bar{z}\}[a,0,\bar{c}][0,\bar{e},0]$ |
| 4 | h | $.2^*.$ | $\{0,y,\frac{1}{2}\}[0,b,0][d,0,f]$ | | $\{\frac{1}{2},\bar{y},\frac{1}{2}\}[0,\bar{b},0][\bar{d},0,f]$ |
| | | | $\{0,\bar{y},\frac{1}{2}\}[0,b,0][d,0,f]$ | | $\{\frac{1}{2},y,\frac{1}{2}\}[0,\bar{b},0][\bar{d},0,f]$ |
| 4 | g | $.2^*.$ | $\{0,y,0\}[0,b,0][d,0,f]$ | | $\{\frac{1}{2},\bar{y},0\}[0,\bar{b},0][\bar{d},0,f]$ |
| | | | $\{0,\bar{y},0\}[0,b,0][d,0,f]$ | | $\{\frac{1}{2},y,0\}[0,\bar{b},0][\bar{d},0,f]$ |
| 2 | f | $mm'^*2'^*$ | $\{\frac{1}{4},\frac{1}{2},z\}[a,0,0][d,0,0]$ | | $\{\frac{3}{4},\frac{1}{2},\bar{z}\}[\bar{a},0,0][\bar{d},0,0]$ |
| 2 | e | $mm'2'$ | $\{\frac{1}{4},0,z\}[a,0,0][0,0,0]$ | | $\{\frac{3}{4},0,\bar{z}\}[\bar{a},0,0][0,0,0]$ |
| 2 | d | $.2^*/m'^*.$ | $\{0,\frac{1}{2},\frac{1}{2}\}[0,0,0][d,0,f]$ | | $\{\frac{1}{2},\frac{1}{2},\frac{1}{2}\}[0,0,0][\bar{d},0,f]$ |
| 2 | c | $.2^*/m'.$ | $\{0,0,\frac{1}{2}\}[0,0,0][0,0,0]$ | | $\{\frac{1}{2},0,\frac{1}{2}\}[0,0,0][0,0,0]$ |
| 2 | b | $.2^*/m'^*.$ | $\{0,\frac{1}{2},0\}[0,0,0][d,0,f]$ | | $\{\frac{1}{2},\frac{1}{2},0\}[0,0,0][\bar{d},0,f]$ |
| 2 | a | $.2^*/m'.$ | $\{0,0,0\}[0,0,0][0,0,0]$ | | $\{\frac{1}{2},0,0\}[0,0,0][0,0,0]$ |

**Figure 1** Table of crystallographic properties of the double antisymmetry space group P(1,1*,1) mm'a*, No. 10977.

## 2. Tables of crystallographic properties of double antisymmetry space groups

The tables contain crystallographic properties for all 17,803 double antisymmetry space group types. Each table contains:

(1) Headline

(2) Lattice diagram

(3) Diagram of symmetry elements and general position diagram

(4) Symmetry operations

(5) General and special positions.

### 2.1. Headline

A headline is placed on the left top of the first page for each table. Each headline consists of two lines, which read from left to right. A headline consists of the following information:

*First line*:

(1) The serial number of the double antisymmetry space group

The serial numbers of the double antisymmetry space groups follow the numbering system of VanLeeuwen et al (2013)

(2) International (Hermann-Maugin) symbol of the double antisymmetry space group.

The International symbols are those introduced by VanLeeuwen et al (2013). The notation specifies the representative group of the group type. The symmetry operations of the representative group are given later in the section of symmetry operations.

(3) X-ray diffraction symmetry group.

This is the symmetry which would be expected to be indicated by the typical methods of analysis of x-ray diffraction patterns (one of the 230 conventional space groups). This group comes from converting all primed operations to colorless operations (because x-ray scattering is invariant under the action of time-reversal) and removing all starred and primed-starred operations (because x-ray scattering is non-invariant under the action of rotation-reversal). If the standard setting of the X-ray diffraction symmetry group differs from that of the double antisymmetry space group, the affine transformation relating the two is given. The notation for this transformation is meant to be short-hand for the 4x4 augmented matrix of the transformation, i.e.

$$\begin{pmatrix} R_{11} & R_{12} & R_{13} & t_1 \\ R_{21} & R_{22} & R_{23} & t_2 \\ R_{31} & R_{32} & R_{33} & t_3 \\ 0 & 0 & 0 & 1 \end{pmatrix}$$

would be given as $R_{11}a + R_{12}b + R_{13}c + t_1, R_{21}a + R_{22}b + R_{23}c + t_2, R_{31}a + R_{32}b + R_{33}c + t_3$.

*Second line:*

(1) The serial number of the colorblind parent group type in the *International Tables for Crystallograph*, Vol. A (2006) (abbreviated as ITC-A). The colorblind parent group of a double antisymmetry space group is the space group based on which the double antisymmetry space group is generated. It is denoted by **Q** by VanLeeuwen et al. (VanLeeuwen *et al.*, 2013) The colorblind parent group of a double antisymmetry space group can be derived from the double antisymmetry space group by substituting all double identities coupled with the symmetry operations with identities.

(2) double antisymmetry point group symbol to which the space group belongs

(3) The name of the crystal system of the group

For example, the headline for No. 10977 from Fig. 1 is shown below.

|  |  |  |
|---|---|---|
| No. 10977 | P(1,1*,1) mm'a* | Pmmn |
|  |  | a, $2b + \frac{1}{2}$, c |
| SG. 51 | mm'm1* | Orthorhombic |

The serial number of the group type is 10977, the modified International (Hermann-Maugin) symbol is P(1,1*,1) mm'a*, the colorblind parent group is space group No. 51 in ITC-A, the double antisymmetry point group in which the group belongs is mm'm1*, and the crystal system of the group is orthorhombic.

**2.2. Lattice diagram**

A three-dimensional lattice diagram is given for each double antisymmetry space group in the upper right corner of the first page, which describes the conventional lattice cell of the colorblind parent group of the group with the generators of the translation subgroup of the group. For example, the lattice diagram of group P(1,1*,1) mm'a* is shown in Figure 2. The conventional lattice cell of the group type is orthorhombic and the generators of the translation subgroup are (1, 0, 0), (0, 1, 0)*, and (0, 0, 1). The color of each arrow indicates whether it is a pure colorless translation (black), a primed translation (red), a starred translation (blue), or a prime-starred translation (green). For instance, the blue arrow along y in Fig. 2 represents a starred translation along y.

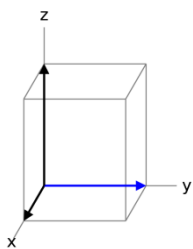

**Figure 2** Lattice diagram of double antisymmetry space group P(1,1*,1)mm'a*, No. 10977. The diagram shows the orthorhombic P(1,1*,1) lattice. The unit translation along y-axis is colored blue, which shows the translation is coupled with the anti-identity 1*.

**2.3. Diagrams of symmetry elements and general positions**

Following the headline and the lattice diagram, diagrams of symmetry elements and general positions are given. The symmetry element diagrams show the relative locations and orientations of the symmetry elements and the general position diagrams show locations of a set of general positions.

The arrangement of the diagrams for different crystal systems are shown in Table 3. The projection direction for all diagrams is perpendicular to the plane of the figure. The *b*-axis is selected as the unique axis for monoclinic groups. Symmetry element diagrams projected along *a*, *b*, and *c* directions are given for triclinic, monoclinic and orthorhombic groups. Symmetry element diagrams projected along the c-axis are given for tetragonal, hexagonal, and trigonal groups. General position diagrams projected along c-axis are given for all groups.

The color of a graphical symbol of a symmetry element indicates that the symmetry element is colored with an anti-identity associated with that color or not coupled with any anti-identity if the color is black. Similarly, the color of a graphical symbol of a general position indicates that the position is generated from the starting position by a symmetry operation coupled with anti-identity associated with that color or not coupled with any anti-identity if the color is black. The colors associated with antisymmetry identities follow the coloring schemes of VanLeeuwen et al (2013) which is listed in Table 1.

Two types of unit cell are defined for a double antisymmetry space group. Here we define the conventional unit cell of the colorblind parent group of a double antisymmetry space group as the *colorblind unit cell*, and the unit cell that can fill the space with colorless translations as the *colorless unit cell*. A colorless translation is a translation that is not coupled with any anti-identity. A colored translation is a translation coupled with one of the three anti-identities, and is printed in the same color as the anti-identity to denote the coupling. The colorblind unit cell and the colorless unit cell are different for double antisymmetry space groups in which the colorblind unit cell translations are coupled with anti-identities. For example, in a P(1,1*,1) lattice, the colorblind unit cell is the same as its colorblind parent space group, which has a size of 1x1x1, while the colorless unit cell has a size of 1x2x1.

The diagrams only consider the unit cell of the colorblind parent group of a double antisymmetry space group, i.e. the colorblind unit cell. The diagrams only show the symmetry elements within the colorblind unit cell, because the diagrams can be easily extended as the symmetry element diagrams are periodic in the directions of translation, and general position diagrams are periodic in the direction of a colorless translation and the color alternating according to color operation rule in the direction of a colored translation.

### 2.3.1. Diagrams of symmetry elements

The symbols used in diagrams of symmetry elements are extensions of those used in ITC-A and are listed in Table 2. Detailed meanings of the symbols are discussed in Appendix B.

The heights of centers of symmetry, rotoinversions, and axes and planes parallel to the plane of projection are printed next to the graphical symbol if the heights are non-zero. Some symmetry elements sit on top of each other in the projected diagram. The way heights are provided for these elements are listed in Table 10.

If a group has a colored unit cell translation and there is a mirror plane which contains the translation, combination of the translation and the mirror operation will result in a glide plane with a glide vector of colorblind unit cell translation. Such glide vectors are not shown in the diagrams as they can be derived from the original two operations. Because these two operations share the same plane showing both would unnecessarily complicate these diagrams. This also applies to any screw axes that come from a rotation axes coupled with a colored unit cell translation. The main benefit of this way of presenting the symmetry element diagrams is that the diagram of a double antisymmetry space group will show exactly the same elements as symmetry element diagram of its colorblind parent groups, with only additional colors to indicate the coupled anti-identities.

### 2.3.2. Diagrams of general positions

The symbols used in diagrams of general positions are extensions of those used in ITC-A. Each position is represented by a circle colored with the color of the position and a height notation beside the circle indicating the z-component of the position. For positions with z-component of "+z" or "-z", the height notation is "+" or "-" respectively. For positions with z-component of "h+z" or "h-z", the height notation is "h+" or "h-" respectively. If two general positions have the same x and y component and z components of +z and –z respectively, the two positions are represented as 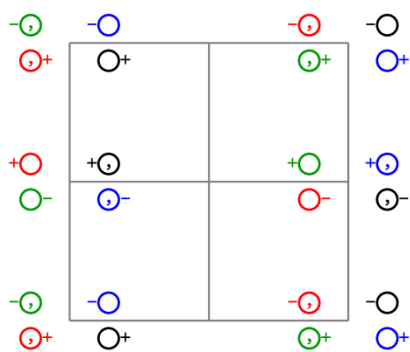. Each half of the symbol correspond to a unique position. The height notations are placed on two sides of the symbol respectively. For positions with a reversed chirality with respect to the starting position, a comma is added into the circle or the half of the circle that represents that position.

**Figure 3** General position diagram for the double antisymmetry space group P(1,1*,1) mm'a*, No. 10977.

### 2.4. Symmetry operations

The modified ITC-A notation and Seitz notation for the symmetry operations are listed under the heading *Symmetry operation*. The symbolism follows Section 11.1.2 of ITC-A with the addition of the use of prime, star, and prime-star to denote coupled anti-identities. For glide planes and screw axes, the glide and screw vectors are given in unit of the length of the colorblind unit cell. In addition, a Seitz notation is also given for each symmetry operation.

For double antisymmetry space groups with centred cells, the symmetry operations are given in several blocks. Groups without centering translations only have one block. For groups with centering translations, the number of blocks in addition to the origin (0, 0, 0)+ block equals to the number of the centring translations. For example, in a group with colorless body centering translations, two blocks (0, 0, 0)+ and (1/2, 1/2, 1/2)+ are shown. The transformations of symmetry operation are explicitly given in the notations, so readers do not need to add these centring translations to the notations.

## 2.5. General and special positions with spin vectors and roto vectors

The position table of a double antisymmetry space group under *Positions* consists of general positions and special positions. These positions are called Wyckoff positions. The first block of the table is the general positions, which are the points that are left invariant only by identity operation or colored identity operations. The remaining blocks of the table are the special positions, which are the points that are left invariant by at least one non-identity operation. For each entry, the columns contains the following information from left to right.

      a. Multiplicity of the Wyckoff position

      b. Wyckoff Letter

      c. Oriented site-symmetry symbol

      d. Coordinates and vectors on sites

### 2.5.1. Multiplicity

The multiplicity is the number of equivalent positions in the conventional unit cell of the colorblind parent space group of the double antisymmetry space group.

### 2.5.2. Wyckoff Letter

The letter is a coding scheme for the blocks of positions, starting with *a* at the bottom block and continuing upwards in alphabetical order.

### 2.5.3. Oriented site-symmetry symbol

The site symmetry group of the first position of each block of positions is given by an oriented symbol. The group is isomorphic to a subgroup of the point group of the double antisymmetry space group. The symbol shows how the symmetry elements of the site symmetry group are oriented at the site. The symmetry element symbol was placed according to the sequence of symmetry directions in the space group symbol. The symmetry directions that do not contribute any element to the site symmetry are represented by dots.

### 2.5.4. Coordinates and vectors on sites

For each block of positions, the coordinates of the positions are given. Immediately to the right of each set of coordinates are two sets of components of two types of symmetry-allowed vectors at that position. The types of vectors are spin vector and roto vector placed on the left and right respectively. The properties of the two types of vectors are listed in Table 2. Then the components of the two types of vectors at the other positions are determined by applying the symmetry operations to the components of the corresponding vector at the first position.

**Table 2** Properties of two types of vectors listed in Wyckoff position table

| Type of vector | Axial or polar | Inverted by 1' | Inverted by 1* |
|---|---|---|---|
| Spin vector | axial | Yes | No |
| Roto vector | axial | No | Yes |

For double antisymmetry space groups with centred cells, the centring translations such as (0,0,0)+ (1/2,1/2,1/2)'+ are listed above the ordinate triplets. The symbol "+" means that the components of these centering translation coordinates should be added to the listed coordinate triplets.

### Acknowledgements

We acknowledge support from the Penn State Center for Nanoscale Science through the NSF-MRSEC DMR #0820404. We also acknowledge NSF DMR-0908718 and DMR-1210588.

**Appendix A. Schematic representations of diagram arrangement for different crystal systems**

**Table 3**   Schematic representations of the general-positions and symmetry-elements diagrams for different crystal systems (**G** = general position diagram)

Triclinic:

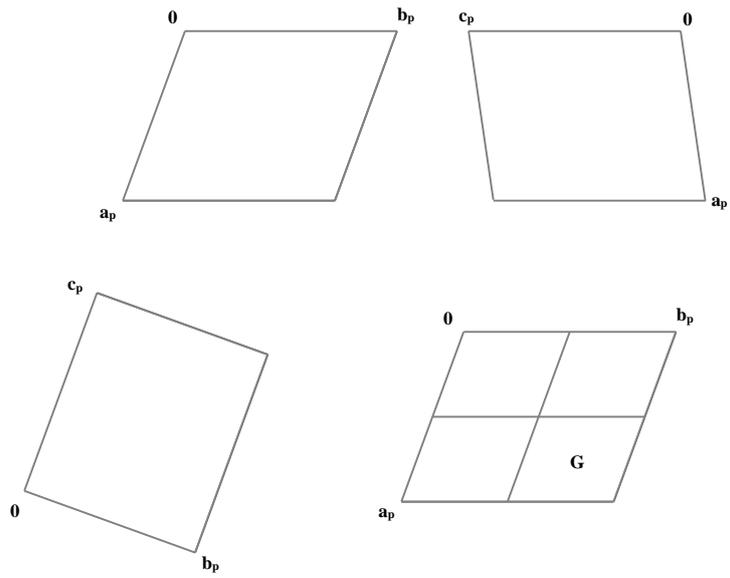

Monoclinic:

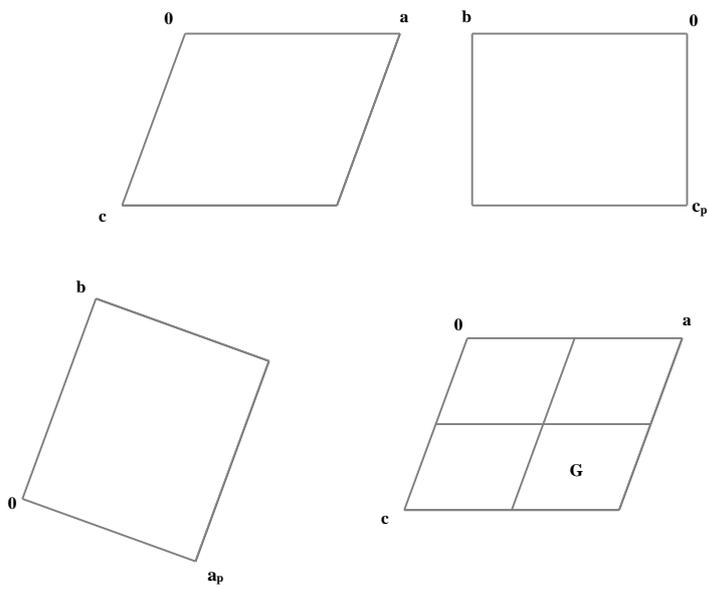

Orthorhombic:

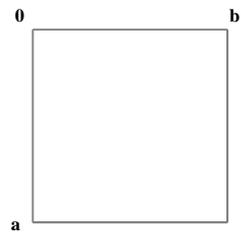 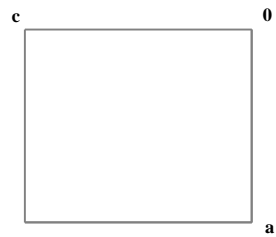

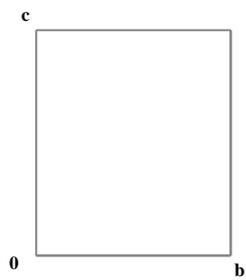 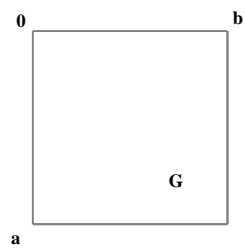

Tetragonal:

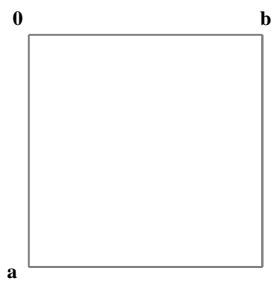 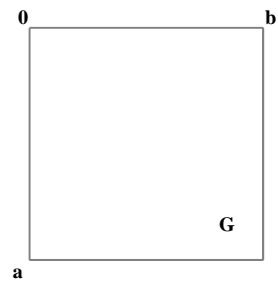

Trigonal, Hexagonal and Rhombohedral:

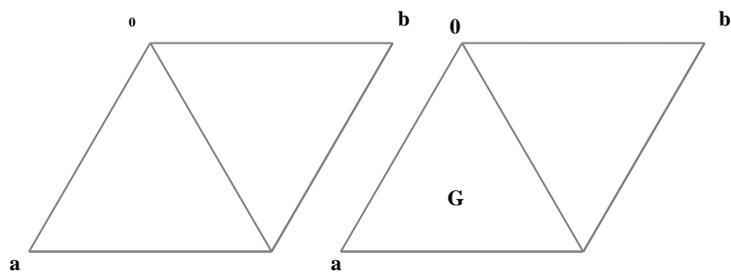

**Appendix B. Double antisymmetry space group diagram symbols**

Most symbols used in the double antisymmetry space group diagrams have the same shapes as those used in ITC-A. The symbols are colored according to the anti-identity coupled with the operation. Extra symbols come from colored unit cell translations on the projection direction of the diagrams. When the unit cell translation is coupled with an anti-identity, there will be symmetry elements with different colors that are located at the same position on the projection plane but at different positions along the projection direction. Both the original and the extra symbols used are listed in the following tables (Table 4-10). The symbols are based on a symmetry elements font *cryst* created by Ulrich Müller from http://www.iucr.org/resources/symmetry-font.

For simplicity, any symmetry element that has only one color associated with it is represented by a symmetry element coupled with 1' (red), in Table 4-10. 1' can be substituted by identity or any anti-identity (1', 1*, or 1'*) and the coloring should be substituted (to black, red, blue, or green) respectively. Similarly, symmetry elements with two colors are represented by symmetry elements with 1' and 1* (red and blue) in Table 4-10 where 1' and 1* can be substituted by any two anti-identities and the colorings should be substituted respectively.

**Table 4**  Symmetry axes parallel to the plane of projection

| Symmetry axis | Graphical symbol | Screw vector of a right-handed screw axis in units of colorblind unit cell translation parallel to the axis | Printed symbol |
|---|---|---|---|
| 2-fold primed rotation axis | | None | 2' |
| 2-fold primed screw axis | | 1/2 | $2_1$' |
| 4-fold primed rotation axis | | None | 4' |
| 4-fold primed screw axis | | 1/4 | $4_1$' |
| 4-fold primed screw axis | | 2/4 | $4_2$' |

| Symmetry axis or symmetry point | Graphical symbol | Screw vector of a right-handed screw axis in units of colorblind unit cell translation parallel to the axis | Printed symbol |
|---|---|---|---|
| 4-fold primed screw axis | | 3/4 | $4_3'$ |
| 4-bar primed inversion | | None | $\bar{4}'$ |

Table 5  Symmetry axes normal to the plane of projection

| Symmetry axis or symmetry point | Graphical symbol | Screw vector of a right-handed screw axis in units of colorblind unit cell translation parallel to the axis | Printed symbol |
|---|---|---|---|
| Identity | None | None | 1 |
| Primed center of symmetry | ○ | None | $\bar{1}'$ |
| Primed center of symmetry and starred center of symmetry | ⊙ | None | $\bar{1}'$, $\bar{1}*$ |
| 2-fold primed rotation axis | | None | $2'$ |
| 2-fold primed screw axis | | 1/2 | $2_1'$ |
| 3-fold primed rotation axis | ▲ | None | $3'$ |
| 3-fold primed screw axis | | 1/3 | $3_1'$ |
| 3-fold primed screw axis | | 2/3 | $3_2'$ |
| 4-fold primed rotation axis | ◆ | None | $4'$ |
| 4-fold primed screw axis | | 1/4 | $4_1'$ |
| 4-fold primed screw axis | 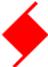 | 2/4 | $4_2'$ |
| 4-fold primed screw axis | 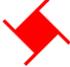 | 3/4 | $4_3'$ |
| 4-fold primed inversion | 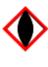 | None | $\bar{4}'$ |
| 6-fold primed rotation axis | 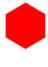 | None | $6'$ |
| 6-fold primed screw axis | 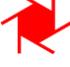 | 1/6 | $6_1'$ |
| 6-fold primed screw axis | 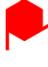 | 2/6 | $6_2'$ |
| 6-fold primed screw axis | 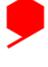 | 3/6 | $6_3'$ |
| 6-fold primed screw axis | 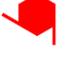 | 4/6 | $6_4'$ |
| 6-fold primed screw axis | 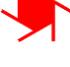 | 5/6 | $6_5'$ |
| 6-bar primed inversion | 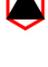 | None | $\bar{6}'$ |

If a center of symmetry is on a rotation axes, the symbol of the center of symmetry is placed onto the symbol of the rotation axis.

**Table 6** Symmetry planes normal to the plane of projection

| Symmetry plane | Graphical symbol | Glide vector in units of colorblind unit cell translation parallel to the projection plane | Printed symbol |
|---|---|---|---|
| Primed mirror plane | 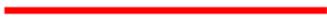 | None | m' |
| Primed glide plane | 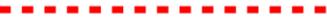 | 1/2 lattice vector along line in projection plane | a', b', or c' |

| Primed glide plane | 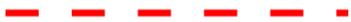 | 1/2 lattice vector normal to projection plane | a', b', or c' |
|---|---|---|---|
| Primed glide plane | 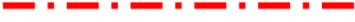 | One glide vector with two components: 1/2 along line parallel to projection plane, 1/2 normal to projection plane | n' |
| Primed glide plane and starred glide plane | 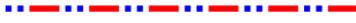 | Two glide vectors: 1/2 along line parallel to projection plane primed and 1/2 normal to projection plane starred | e1' and e2* |
| Primed glide plane | 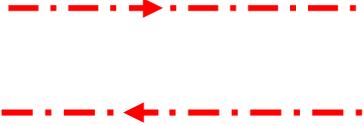 | 1/4 along line parallel to projection plane, combined with 1/4 normal to projection plane (arrow indicates direction parallel to the projection plane for which the normal component is positive) | d' |

**Table 7**  Symmetry planes parallel to the plane of projection

| Symmetry plane | Graphical symbol | Glide vector in units of colorblind unit cell translation parallel to the projection plane | Printed symbol |
|---|---|---|---|
| Primed mirror plane | 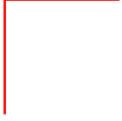 | None | m' |
| Primed glide plane | 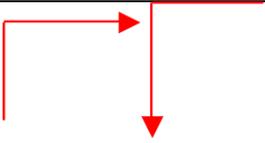 | 1/2 | a', b', or c' |

| Primed glide plane | 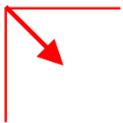 | One glide vector with two components, 1/2 in the direction of the arrow | n' |
| --- | --- | --- | --- |
| Primed glide plane and starred glide plane | 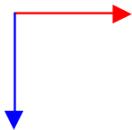 | Two glide vectors, 1/2 in either of the directions of the two arrows | e1' and e2* |

**Table 8** Symmetry axes inclined to the plane of projection (cubic groups only)

| Symmetry axis | Graphical symbol | Screw vector of a right-handed screw axis in units of colorblind unit cell translation parallel to the axis | Printed symbol |
| --- | --- | --- | --- |
| 2-fold primed rotation axis | 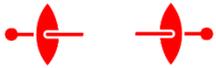 | None | 2' |
| 2-fold primed screw axis | 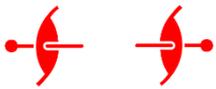 | 1/2 | $2_1$' |
| 3-fold primed rotation axis | 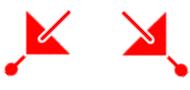 | None | 3' |
| 3-fold primed screw axis | 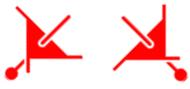 | 1/3 | $3_1$' |
| 3-fold primed screw axis | 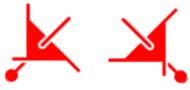 | 2/3 | $3_2$' |

**Table 9** Symmetry planes inclined to the plane of projection (cubic groups only)

| Symmetry plane | Graphical symbol for planes normal to | | Glide vector in units of colorblind unit cell translation for planes normal to | | Printed symbol |
| --- | --- | --- | --- | --- | --- |
| | [011] and [01$\bar{1}$] | [101] and [10$\bar{1}$] | [011] and [01$\bar{1}$] | [101] and [10$\bar{1}$] | |
| Primed mirror plane | 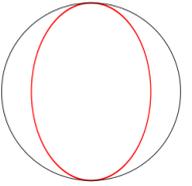 | 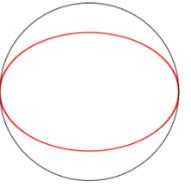 | None | None | m' |
| Primed glide plane | 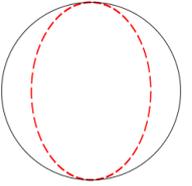 | 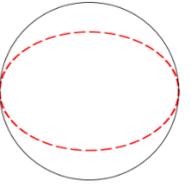 | 1/2 lattice vector along [100] | 1/2 lattice vector along [010] | a' or b' |
| Primed glide plane | 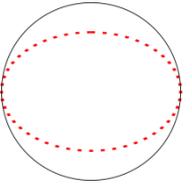 | 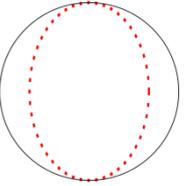 | 1/2 lattice vector along [01$\bar{1}$] or along [011] | 1/2 lattice vector along [10$\bar{1}$] or along [101] | a' or b' |
| Primed glide plane And starred glide plane | 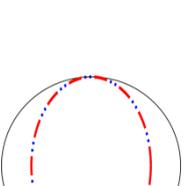 | 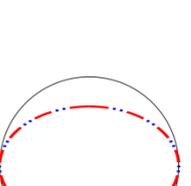 | Two glide vectors: 1/2 along [100] primed and 1/2 along [01$\bar{1}$] or along [011] starred | Two glide vectors: 1/2 along [010] primed and 1/2 along [10$\bar{1}$] or along [101] starred | e1' and e2* |
| Primed glide plane | 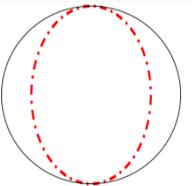 | 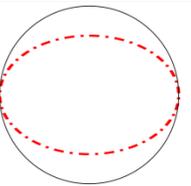 | One glide vector: 1/2 along [11$\bar{1}$] or along [111] | One glide vector: 1/2 along [11$\bar{1}$] or along [111] | n' |
| Primed glide plane | 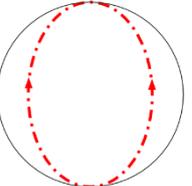 | 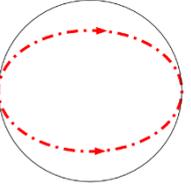 | 1/2 along [1$\bar{1}$1] or along [111] | 1/2 along [$\bar{1}$11] or along [111] | d' |
| Primed glide plane | 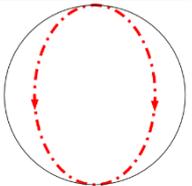 | 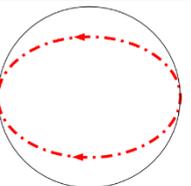 | 1/2 along [$\bar{1}\bar{1}$1] or along [$\bar{1}$11] | 1/2 along [$\bar{1}\bar{1}$1] or along [1$\bar{1}$1] | d' |

**Table 10** Heights of symmetry operations above plane of projection

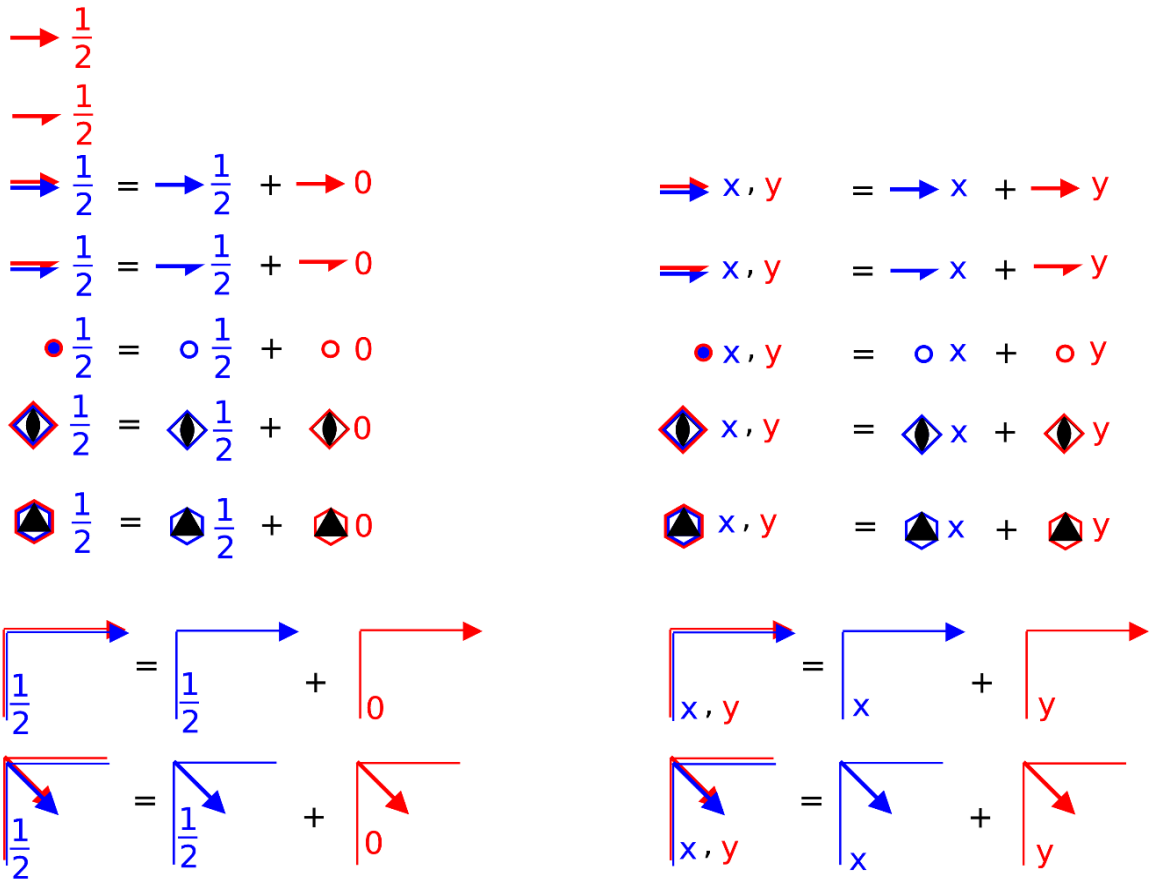

The texts besides the symbols indicate the height of the symbols. *x* and *y* represent arbitrary heights.